# Photonic Advantage of Optical Encoders

**Authors:** Luocheng Huang[1], Quentin A. A. Tanguy[1], Johannes E. Fröch[1,2], Saswata Mukherjee[1], Karl F. Böhringer[1,3,4], Arka Majumdar[1,2, *]

**Affiliations:**

[1]Electrical and Computer Engineering, University of Washington, Seattle, WA-98195

[2]Physics Department, University of Washington, Seattle, WA-98195

[3]Department of Bioengineering, University of Washington, Seattle, WA, 98195, USA

[4]Institute for Nano-Engineered Systems, University of Washington, Seattle, WA, 98195, USA

* Corresponding author. Email: arka@uw.edu

**Abstract:**

Light's ability to perform massive linear operations parallelly has recently inspired numerous demonstrations of optics-assisted artificial neural networks (ANN). However, a clear advantage of optics over purely digital ANN in a system-level has not yet been established. While linear operations can indeed be optically performed very efficiently, the lack of nonlinearity and signal regeneration require high-power, low-latency signal transduction between optics and electronics. Additionally, a large power is needed for the lasers and photodetectors, which are often neglected in the calculation of energy consumption. Here, instead of mapping traditional digital operations to optics, we co-optimized a hybrid optical-digital ANN, that operates on incoherent light, and thus amenable to operations under ambient light. Keeping the latency and power constant between purely digital ANN and hybrid optical-digital ANN, we identified a low-power/ latency regime, where an optical encoder provides higher classification accuracy than a purely digital ANN. However, in that regime, the overall classification accuracy is lower than what is achievable with higher power and latency. Our results indicate that optics can be advantageous over digital ANN in applications, where the overall performance of the ANN can be relaxed to prioritize lower power and latency.

**One-Sentence Summary:** We clearly demonstrate superior performance of a co-designed optical encoder and digital backend over a purely digital artificial neural network under constraints of low-latency and low-power.

**Main Text:**

**Introduction:**

Over the last decade the fields of artificial intelligence (AI) and deep learning have experienced accelerated progress, revealing the potential and capabilities of artificial neural networks (ANN) for a variety of applications, with recent demonstrations even advancing to the public spotlight in the form of chat software and artistic rendering programs. Their recent success can be traced back to major breakthroughs, both in terms of computational algorithms and digital hardware such as graphics processing units (GPU) (*1*). While impressive, the scaling of power and latency of digital implementations of deep learning turned out to be unfavorable with the size of the ANN. This poses a serious limitation for further scaling of ANN (*2, 3*) and applicability to low-power, real-time problems.

Light may be the answer to this scaling, thanks to its inherent parallelism, speed, and analog nature, thus providing an attractive alternative to electronic implementations to build energy efficient and fast artificial neural networks (ANNs). This has been recognized early on and several experiments reported optical ANNs already back in the 1990s (*4, 5*). Unfortunately, progress stalled due to technological and fundamental reasons, which can be broadly classified into intrinsic and extrinsic problems. Intrinsic problems with optics had been the large size and poor tolerance to misalignment of optical components; limited space bandwidth product of spatial light modulators; and lack of nonlinear activation. The extrinsic problems originated from poor understanding of AI algorithms and adaptive learning, as well as the meteoric rise of electronic computing systems.

Given the current limitations of electronic hardware and our increased understanding of AI, the extrinsic problems are somewhat alleviated. In parallel, the advancement of nano-fabrication facilities, and the availability of sophisticated electromagnetic simulators have led to the high-volume manufacturing of multi-functional nano-optics, such as flat meta-optics (*6, 7*) and integrated photonic devices (*8*). Emerging material systems coupled with these nano-optical structures enable monolithic photonic integrated circuits (PIC) analogous to electronic ICs (*9*). These innovations in nanophotonics and AI, combined with severe limitations of digital implementation of ANNs have generated strong interest in recent years in recreating optics assisted ANNs (*10-17*).

However, thus far, none of the reported works has demonstrated a clear advantage of optics over digital ANNs for inference. Most implementations have only shown the substitution of a small linear part with an optical counterpart, while the rest was kept in the digital electronics. Although there is a clear advantage of optics for implementing a small sub-system, often the linear part, the power and latency in a complete ANN include the transduction of the signal between optical and electronic domains (*18*), i.e. the detector readout power and laser power, many of which are often neglected. In fact, an analysis considering these energy costs shows that implementing only one convolutional layer in optics does not provide any advantage, unless the input has a very large dimension (*18*). Additionally, a large body of works demonstrated classification for extremely simple "toy" problems, for which no digital benchmark exists (*13, 14*). Comparing the power and latency of an application specific optical ANN to a GPU (optimized for universal operations) is unfair. There are many ways to drastically reduce the power and latency of a digital ANN, including replacing matrix multiplication with XNOR operations (*19*). Many pruning algorithms also exist to reduce the number of computations needed for inference. As such, there has been no clear demonstration where an optics-assisted ANN shows an advantage over a purely digital framework optimized for solving a specific problem. One challenge is that it is impossible to exactly define the computational complexity of an ANN, hence the exact calculation of power and latency in the digital part is dependent on both training and technology.

Here, we develop a framework to exactly compare the inference performance of a pure digital ANN against a hybrid optical-digital ANN. In both ANNs, we ensure the same power and latency, and thus by comparing the classification accuracy, we can clearly assess the relative advantage. Fig. 1 shows the schematic of the two cases: the pure digital and the hybrid optical-digital. We encode the input in incoherent light, as the optical frontend of the ANN can work with ambient light without incurring any additional energy. In a pure digital case, a lens-based sensor captures an image of an object under incoherent light, then the image is transferred to a digital ANN. For the hybrid case, we use an engineered optic – namely the optical encoder, instead of a lens, that captures the image in a different basis and sends the data to a digital backend. Instead of implementing a digital sub-system, such as convolutional operations in optics, we co-optimize the optical frontend (implemented via a sub-wavelength diffractive meta-optics), along with the digital backend using an end-to-end design framework (detail in the supplementary materials S1, S3)(*20*).

The topology and resources (i.e., the same number of nodes, layers, and nonlinearities) used in the digital ANN are kept the same in both cases, though with different weights and biases. Thus, we ensure that the latency and power consumption in both cases remain identical.

Here, we tested the classification accuracy for MNIST data sets for different values of $N$, which represent the binned size of the image captured in the sensor either via a lens or the optical encoder. As the latency and power increase with the input dimensionality $N$ of the data sent to the digital ANN, we found that classification accuracy increases in both cases, and there is no advantage from an optical frontend for large $N$. However, for smaller $N$, where the system power and latency are also lower, we found an increase in validation accuracy ($\sim 10\%$) with a hybrid optical-digital ANN. We experimentally validated our theoretical model. Our work clearly demonstrates a photonic advantage for ANN inference, albeit such an advantage is observed when overall system performance is lower than the highest achievable performance.

**Results:**

Our digital backend consists of three fully connected layers: $N \times 256$ (input), $256 \times 256$ (hidden) and $256 \times 10$ (output). The first two layers are each followed by a rectified linear unit (ReLU) nonlinearity and the output layer has a sigmoid nonlinearity. For the pure digital case, every image is converted to an $N$-pixel image by averaging the pixels. We chose 8 different $N$ ranging from 1 to 100, to assess the performance of the system with increasing data input. We train the digital network by back-propagating the loss function defined by the cross-entropy between the output and the ground truth. In simulation, we obtained a validation classification accuracy of up to $\sim 98\%$ (detail in the supplementary materials S2). We note that, in prior works, to achieve a similar accuracy with the ***MNIST*** dataset, several layers were used (*17*), which we attribute to inefficient training. For the hybrid case, we model the optical frontend using a sub-wavelength diffractive meta-optics, although any freeform optical surface could suffice for implementation. The fabricated optical frontends with different output dimensionalities are shown in Fig. 2 (b). We train the meta-optics along with a digital backend with the same neural network topology (details in method), following a similar framework used before for imaging (*20*). As expected, we observed an increase in classification accuracy with increasing $N$. We also found that for $N > 8 \times 8$, the digital and hybrid ANN demonstrate identical classification accuracies. However, at a lower value of $N$, the classification accuracy of the hybrid ANN surpasses that of the digital ANN. Example classification confusion matrices are shown in Fig. 3 (a), comparing the experimental validation accuracies between a hybrid and a digital ANN with the same input size, $N = 3 \times 3$. Theoretically, we observe an increase in classification accuracy by up to $\sim 20\%$ when an optical frontend is incorporated. A validation accuracy comparison chart can be seen on Fig. 3 (b). We note that, even with a single data-point sent to the digital backend, we expect to see a higher classification accuracy with our optical frontend. This is because that single input can have 256 different values for an 8-bit precision sensor.

To validate the design, we fabricated the meta-optics (detail in the supplementary S4) and measured their performance experimentally, where we projected images of the MNIST data set using an OLED display in green (detail in the supplementary S5). The incoherent green light passes through the meta-optics, and we capture the data on the sensor with 8-bit precision. We then binned the captured image to create the $N$ data-points that are passed to the digital backend. An experimental sample on Fig. 2 (c) shows the signal processing of the $3 \times 3$ encoder. Due to fabrication imperfections, and misalignments, we retrained the digital backend (with the same topology) using the captured data. Our experiment matches the theory very well for $N \geq 3 \times 3$. We note that the meta-optics optimized for $N = 8 \times 8$ was damaged and we could not collect data on that. At smaller $N$, the deviation from the theory is attributed to experimental noise. While a single point can provide more information to the digital backend, it is corrupted by the quantization noise, undermining the effect of the optical encoder and we obtained a similar classification accuracy, as we would have expected from a pure digital backend.

**Discussion:**

By employing an incoherent light source and a meta-optical frontend, we created a framework, enabling us to compare the performance of a digital ANN to an optics-assisted ANN in the same footing. While keeping the power and latency constant in both cases, we showed that optical encoding does provide more information to the digital backend, resulting in $\sim 10\%$ more classification accuracy in the experiment. While our result is primarily applicable to the MNIST dataset, we believe that it indicates the conditions for which an optical frontend is beneficial to increase the performance of an ANN (more discussion in supplementary S6). Without any constraints on latency and power, one can arbitrarily increase $N$, and always find a digital solution that is better than the hybrid option. One way to rationalize this is that any optical implementation can be modelled digitally and therefore without any constraint a digital solution can be found with an accuracy in the same order of magnitude or higher than its optical counterpart. The higher classification accuracy of optics-assisted ANN in several reports is most likely a manifestation of poor training of the fully digital ANN. However, under the constraints of latency or power, we need to work with an intermediate value of $N$, where the optical frontend can provide a more efficient solution, albeit at overall lower accuracy.

**Acknowledgments:**

**Funding:**

National Science Foundation (NSF-ECCS-2127235)

DARPA (Contract # W31P4Q-21-C-0043)

**Author contributions:**

Conceptualization: AM

Methodology: LH, SM, QT, JF

Investigation: LH, SM, QT, JF

Fabrication: QT

Visualization: LH, QT

Funding acquisition: AM, KB

Project administration: AM, KB

Supervision: AM, KB

Writing – original draft: LH, AM, JF

Writing – review & editing: LH, SM, QT, JF, KB, AM

**Competing interests:** AM and KB are co-founders of Tunoptix, which is commercializing similar technology.

**Data and materials availability:** All data are available in the main text or the supplementary materials.

## Supplementary Materials

Materials and Methods

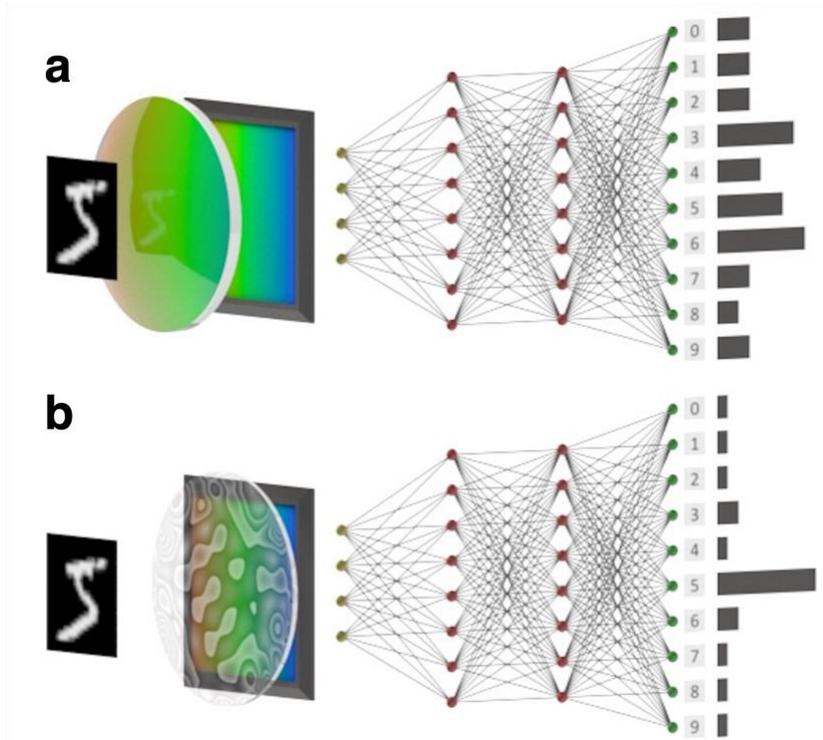

**Fig. 1. Schematic of the optical encoder and pure digital neural network.** (a) Purely digital artificial neural networks operate on captured images in a lensed sensor. (b) Instead of using a lens, a designed optics can perform additional linear operations on the captured data. In both cases, the power and the latency of the sensor are the same. Using the digital computational backend with the same resources (number of layers and neurons), we ensure the same power and latency, both of which monotonically scale with the dimensionality of the input data to the digital backend.

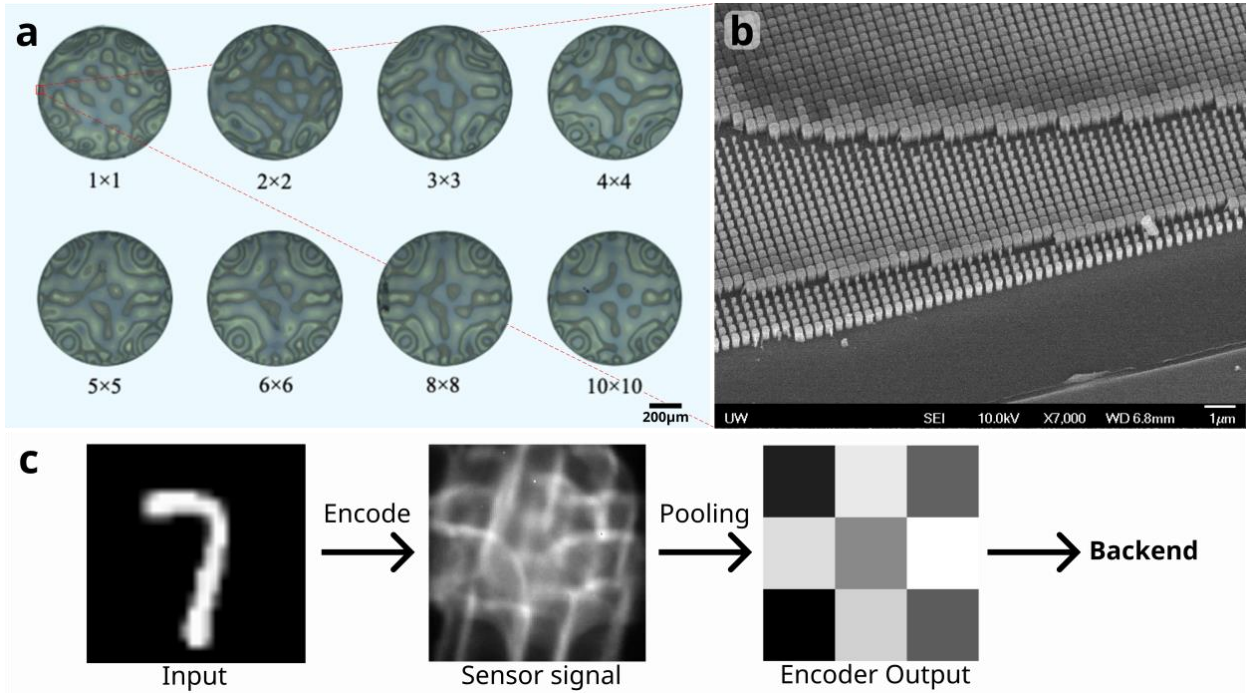

**Fig. 2. Fabrication and characterization of the meta-optical encoder:** (a) Optical microscope images of the meta-optical encoders for different input sizes. (b) Scanning Electron Microscope (SEM) image of the optical encoder, region denoted by the red box on device 1×1. (c) The experimental input, sensor signal, and output of the meta-optical encoder.

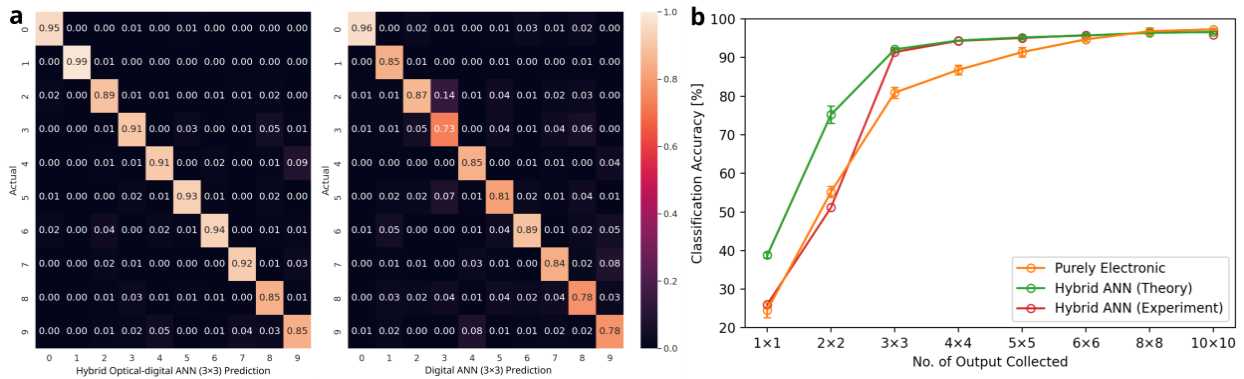

**Fig. 3. Performance comparison of the digital and hybrid ANN.** (a) Confusion matrices comparing the experimental performances of the hybrid optical-digital against the pure digital ANNs for the case of $N = 3 \times 3$. (b) Validation classification accuracies of the purely electronic

and hybrid optical-electronic ANNs. The error bar is shown to represent the range of one standard deviation.

# Supplementary Materials for

## Photonic Advantage of Optical Encoders

Authors: Luocheng Huang[1], Quentin A. A. Tanguy[1], Johannes E. Fröch[1,2], Saswata Mukherjee[1], Karl F. Böhringer[1,3,4], Arka Majumdar[1,2, *]

**This PDF file includes:**

**S1. End-to-end design of the meta-optics and digital backend**

**S2. Design of Purely Digital neural network**

**S3. Design of the meta-optics**

**S4. Fabrication of the meta-optics**

**S5. Meta-optical encoder experimental details**

**S6. The demonstration of benefit of optics in the existing work**

**S1. End-to-end design of the meta-optics and digital backend: the hybrid optical-digital ANN**

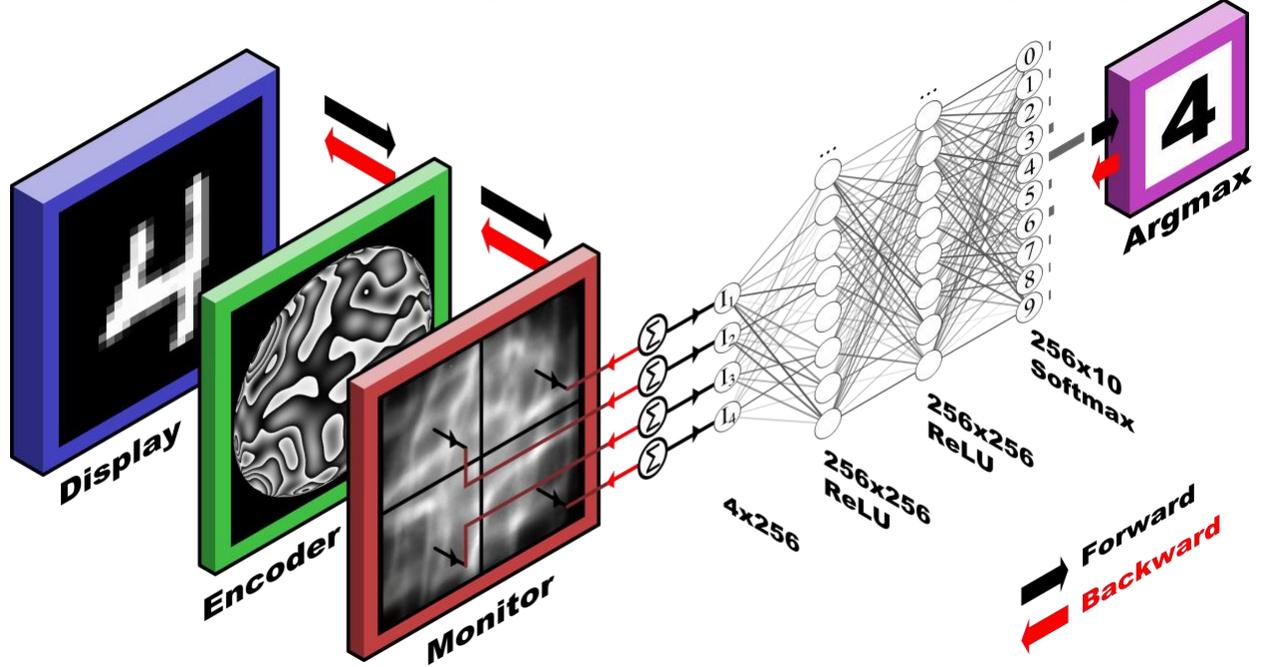

**Figure S1.** The hybrid optical-digital neural network is designed iteratively using an end-to-end differentiable pipeline. Each iteration consists of a forward computation of the loss, and a backward propagation of the loss. An example of the $2 \times 2$ encoding is shown here.

*Design of the Hybrid Optical-digital NN:*

We implemented an end-to-end differentiable pipeline to compute the gradient of the phase distribution on the meta-optical encoder with respect to the classification accuracy of the hybrid optical-digital neural network. The pipeline consists of three stages, namely, point spread function (PSF) simulation, imaging simulation, and classification. The loss function is given by the cross-entropy between the output of the hybrid neural network and the ground truth. We used TensorFlow 2.8 as the automatic differentiation engine to implement the forward computations.

In the PSF simulation stage, light with normal incidence is propagated through the meta-optics, with the phase modulation distribution denoted by $\phi(x, y)$ ("Encoder" in Fig. S1). The phase distribution is parameterized by $z_j$, linear combinations of Zernike polynomials $R_n^m$, in which:

$$R_n^m(\rho) = \sum_{k=0}^{\frac{n-m}{2}} (-1)^k \binom{n-k}{k} \left( \frac{n-2k}{\frac{n-m}{2}-k} \right) \rho^{n-2k}$$

We used 200 terms of the Zernike polynomials to parameterize the phase distribution surface such that $\phi = \sum_{j=1}^{200} z_j R_j$, where $R_j \rightarrow R_n^m$, given $j = \frac{n(n+2)+m}{2}$.

These polynomials are precomputed and stored into the memory, kept until the end of the entirety of the optimization loop. The Zernike polynomials are orthogonal bases that provide spatial regularization to the phase modulation of the metasurface. We found that the employment of such basis functions, instead of optimizing phase value at each spatial location, prevents the optimization from getting stuck in local minima during optimization. After computing the phase distribution, we propagate the complex field using the bandwidth-limited angular spectrum method to obtain the intensity at the focal plane, i.e., the PSF.

The second stage of the end-to-end pipeline is to convolve the PSF with the batched input MNIST images. The input images are first up-sampled using the bilinear interpolation to be the same size as the PSF intensity array. These input images are then stored in memory for the rest of the optimization routine. After that, the PSF is convolved with the input images using the Fourier convolution theorem such that $O = \mathcal{F}^{-1}\{\mathcal{F}\{PSF\} \cdot \mathcal{F}\{I\}\}$, where $O$ denotes the output image, and $I$ denotes the input image. The output $O$ can be seen on the "Monitor" of Fig. S1. We note that this way of modelling imaging is valid only for imaging under incoherent emission, as we aim to demonstrate in this work.

The output images are then fed into a trainable digital backend ANN with sequential layers including an $N \times N$ average pooling layer, two fully connected layers with 256 units each with ReLU activation function, and finally a fully connected layer with 10 units with softmax activation function as the output layer. Note that this digital backend ANN architecture is kept the same between the hybrid optical-digital ANN as well as the purely digital ANN. Note that the $N \times N$ is equal to the input size of the digital backend. While the exact latency and power of the digital backend will depend on the technology (software, GPU or an ASIC optimized for a specific ANN), both of which are expected to monotonically increase with increasing value of $N$.

After the forward computation, we obtain the cross-entropy loss $\mathcal{L}$ between the output of the ANN and the ground truth label of the MNIST data set. The automatic differentiation algorithm then starts the backward computation in which we obtain the gradient of the loss with respect to the Zernike polynomial coefficients, namely $\partial \mathcal{L} / \partial z_j$. This process is visually represented in Fig. S1 by the red arrows. Then we apply this gradient to the parameters $z_j$ multiplied by a factor provided by the Adam optimizer with a learning rate of 0.001.

The training loop includes the forward computation of the loss function, as well as the backward computation of the loss gradient. This training loop is done iteratively until the loss converges. We find that 200 iterations are sufficient for a convergence. Both the forward and backward computations are done on an Intel Xeon @ 2.20 GHz, accelerated by a Nvidia Tesla P100 with 16GB of RAM. The optimization ran 4 times for each input size ($N \times N$).

The training and validation accuracies are displayed on Fig. S2. The training confusion matrices are shown in Fig. S3.

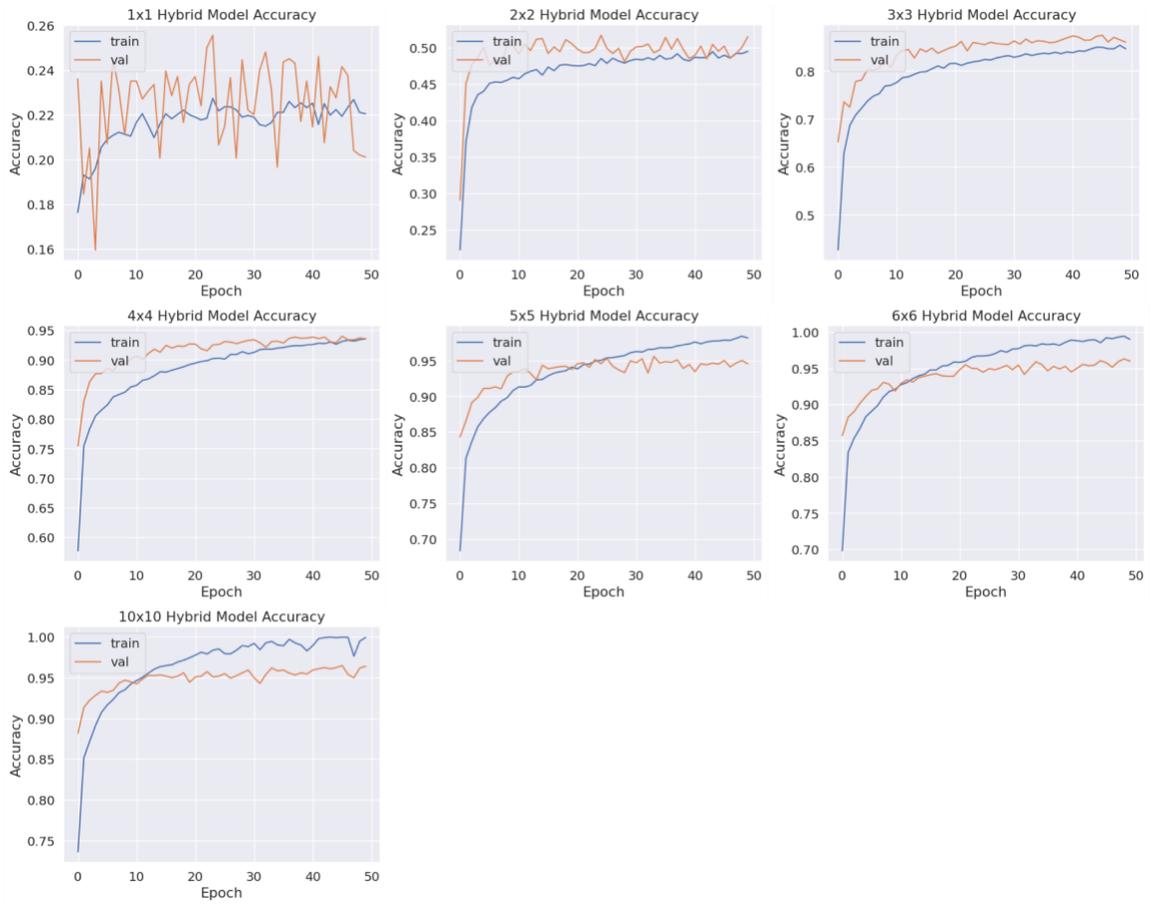

**Figure S2.** The training (train) and validation (val) accuracies of the hybrid artificial neural network classification.

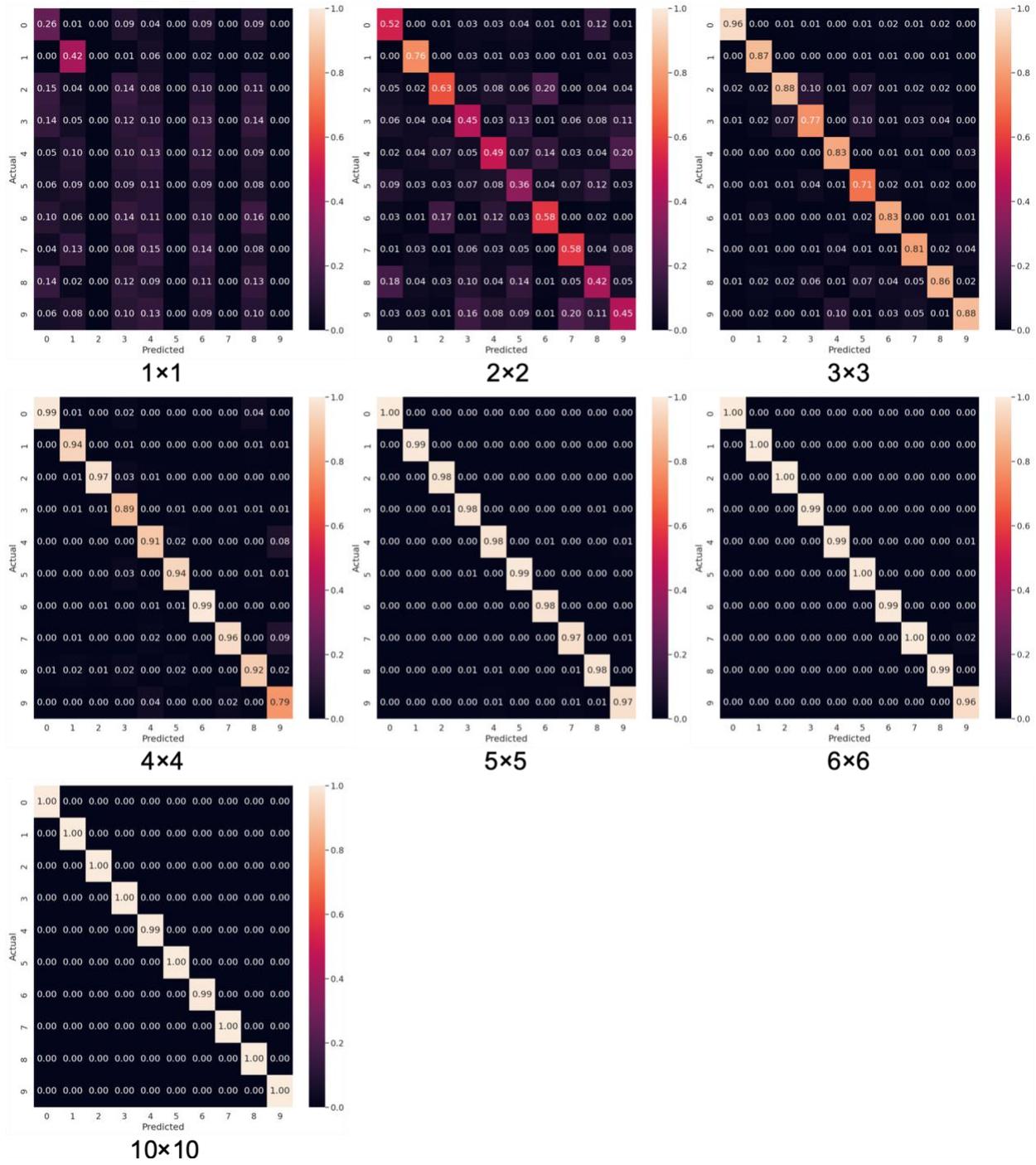

**Figure S3.** The training confusion matrices of the hybrid artificial neural networks.

## S2. Design of Purely Digital neural network

The purely digital artificial neural network is designed with the same sequential architecture as the hybrid neural network's digital backend, comprising four layers: an $N \times N$ average pooling layer; two layers of fully connected neurons with 256 units and ReLU activation functions; and a 10-unit fully connected layer with a softmax activation. The optimization of the neural network follows an identical routine, including the optimizer scheme and the number of iterations, and is conducted using the same hardware. The optimization routine consists of 150 epochs using the Adam optimizer with a learn rate of 0.001. Most of the optimizations converge within ~50 epochs. The training and validation accuracies are displayed on Fig. S4. The training confusion matrices are shown in Fig. S5.

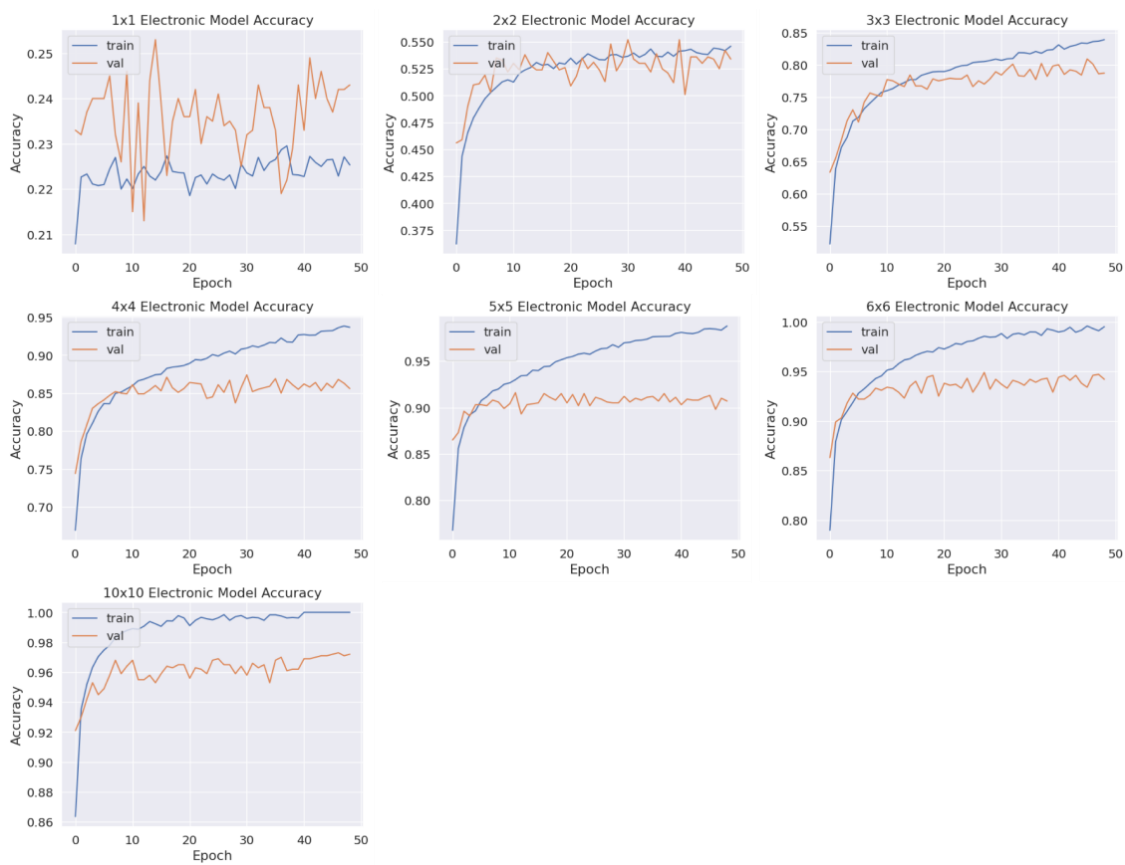

**Figure S4.** The training (train) and validation (val) accuracies of the purely electronic neural networks.

We note that, to ensure the network is well trained, we started with ~5 layers, and reduced the layers and number of neurons and trained with many different inputs. Finally, we achieved a high classification accuracy (~97%) only with two hidden layers. We emphasize that it is important to have a good training of the purely digital ANN, without which we can draw a wrong conclusion on the photonic advantage. We suspect that many of the reported optical neural network works have compared works with a poorly trained digital ANN, showing an improved classification accuracy.

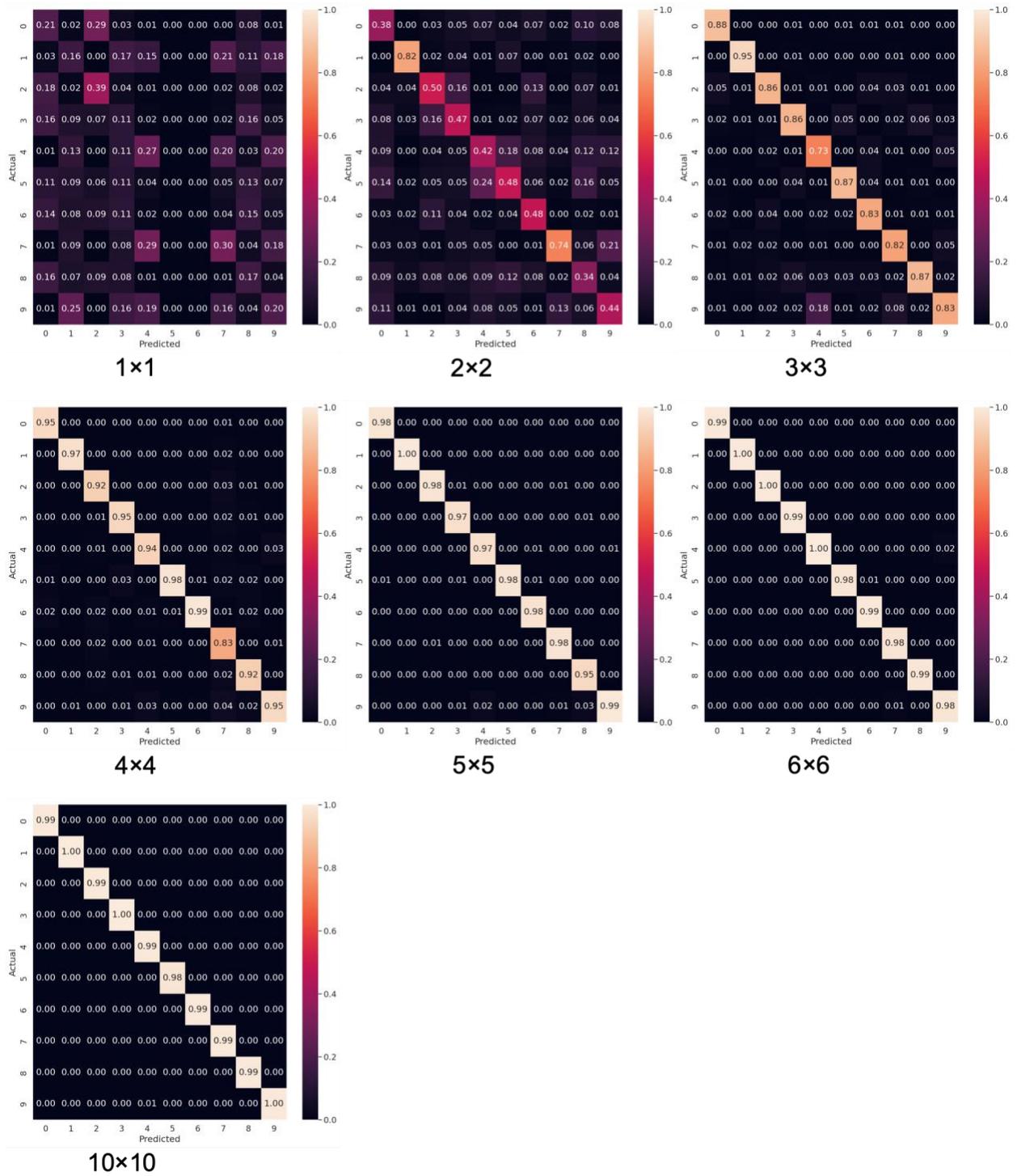

**Figure S5.** The training confusion matrices of the purely digital neural networks.

**S3. Design of the meta-optics**

The metasurface comprises a 2D array of SiN meta-atoms on a Manhattan grid, with a periodicity of 350 nm in both x and y directions. Each meta-atom is shaped as a square pillar, with a fixed height of 775 nm and lateral width parameterized to span the periodicity of the grid. This configuration is simulated using rigorous coupled-wave analysis [1], and the phase modulation of the metasurface is mapped to the corresponding meta-atom that yields the closest phase modulation to the target $\phi(x, y)$. The meta-atoms are situated on a quartz wafer with a thickness of about 500 $\mu m$.

**S4. Fabrication of the meta-optics**

For fabrication, we first deposited a ~ 775 nm thick SiN film on a 500 µm thick quartz wafer using plasma enhanced chemical vapor deposition (PECVD) in a SPTS PECVD chamber. A positive resist (ZEP 520A) was then spun onto the wafer, followed by baking at 180 °C for 3 minutes. To minimize charging effects during patterning, a conductive polymer layer (DisCharge H2O) was subsequently spun on top. The resist layer was then patterned using a 100 kV electron beam (JEOL JBX6300FS) at a dose of ~ 300 µC cm$^{-2}$ and developed in Amyl Acetate for 2 minutes. Then a layer (~ 80 nm) of AlO$_x$ was deposited using electron beam evaporation. After overnight liftoff in NMP heated at 90°C, the SiN layer was etched to the depth of 700 nm (+/-2 nm) with a remaining AlOx thickness of ~10 nm. using a mixture of C$_4$F$_8$/SF$_6$ in an inductively coupled reactive ion etcher (Oxford PlasmaLab System 100). For SEM imaging a thin conductive Au/Pd layer was deposited to prevent charging.

## S5. Meta-optical encoder experimental details

For the hybrid neural network experimental measurements, we first measured the label-feature pairs. The MNIST features, i.e., handwritten black and white images, are scaled up using the nearest neighbor interpolation then displayed on an OLED monitor (SmallHD 5.5 in. Focus OLED HDMI Monitor), as shown on Fig. S6. The image is first displayed via an OLED monitor, set ~10 cm away from the meta-optical encoder. The signal then goes through a custom microscope to transfer the output of the meta-optical encoder to the sensor. The power consumption of the OLED display is ~15 W. Albeit, since the sole purpose of the incoherent light source here is to provide an emulation of the real-world object, the actual power budget of the classification system should not include the OLED monitor.

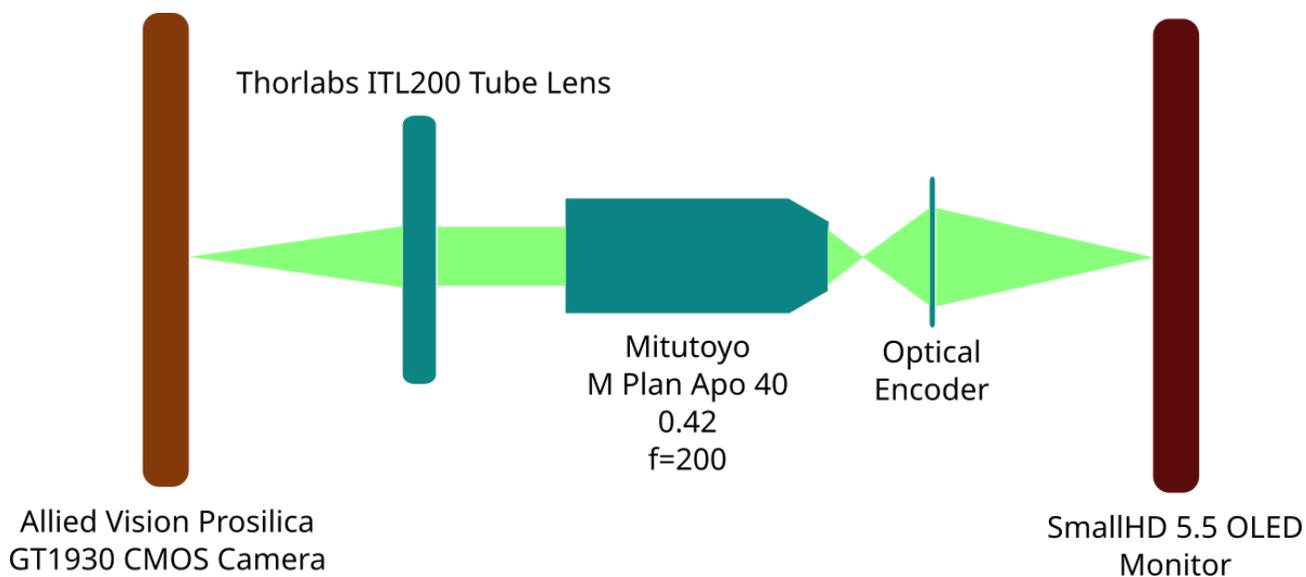

**Figure S6.** Schematic of the experimental setup.

Some examples of the experimental captures are shown in Fig. S8. First, the input is displayed on the OLED monitor shown on the top row. Then the encoder processes the input and projects the signal on the sensor, shown on the second row. The third row shows the encoder output after the average pooling. The fourth row shows what the purely digital ANN receives when there is no optical encoder.

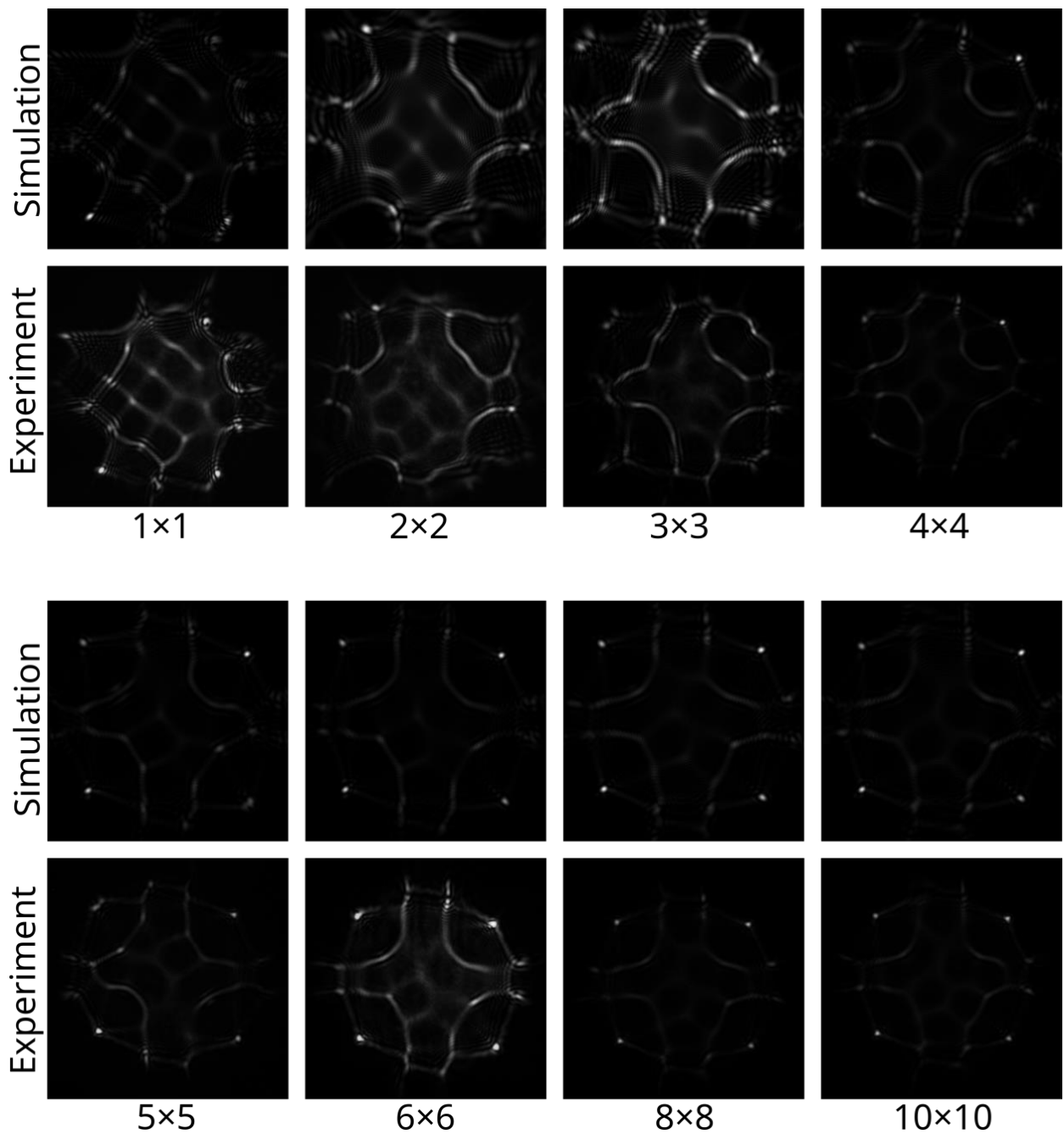

**Figure S7.** Simulation and experimental PSFs of the optical encoders.

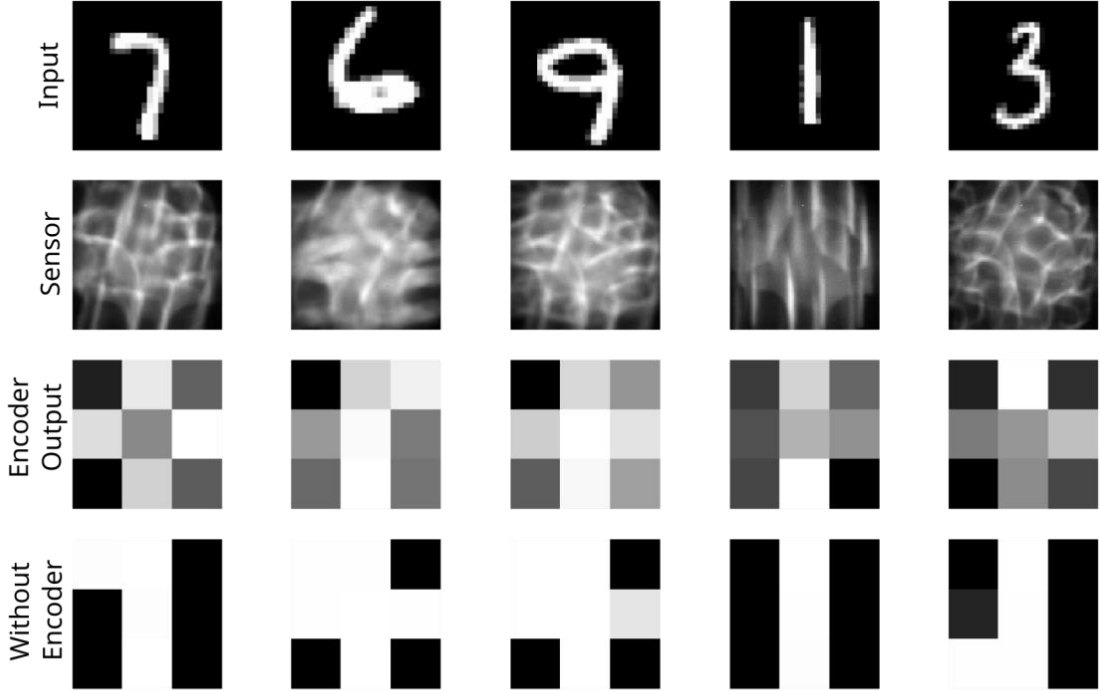

**Figure S8.** Experimental results of the displayed digits, captures, and output of the $3 \times 3$ encoder.

For each meta-optical encoder, we first measured the PSF responses which match up well with the simulated PSFs, seen on Fig. S7. To calibrate the image height on the OLED monitor, we displayed a cross calibration pattern with varying scaling factors and chose the closest image that resembled the simulated output. The images are then judiciously cropped to reflect the correct physical extents of the simulation. The recorded image is then decimated into an $N \times N$ image using average pooling and fed to the digital neural network. Note that one could also implement the average pooling step optically, by using a large pixel size. Such large pixel size can potentially provide benefit for low-light operations, which is important in application in mid/long wave infrared regime.

We captured 10,000 images for the training set and 2,000 images for validation. We used the captured validation data set to first calculate the accuracy of the theoretical hybrid neural network, whose accuracies are shown on Fig. 3(b). The training data set is then used to train a hardware-in-the-loop hybrid ANN with the save topology as the original hybrid ANN. This procedure is done 4 times with different random seeds during training, the validation accuracies of which are recorded. The standard deviation of the validation accuracies is also calculated and shown as the distance between the error bars shown on Fig. 3(b) in the main text.

## S6. The demonstration of benefit of optics in the existing work

While a large body of works exists today on optics-assisted neural network, we argue that none of them showed an advantage over pure digital electronic ANN. Here, our classification of digital electronic ANN encompasses pure software solution (running on a GPU) or an application specific integrated circuit (ASIC) based accelerator. Essentially, as long as we are using CMOS-transistor based hardware, we are classifying them as digital backend. Any nano-electronic solutions, like memristors-based solutions are not included in this analysis. Based on the current works, there are largely three classes of optical neural networks, which are listed as follows.

***Integrated photonics + digital:*** There is an extensive amount of works that rely on integrated photonic based optical operations and digital backends [2]–[7]. Most of them depend on some form of arrays of switches, made of either Mach-Zehnder interferometers or ring resonators. Most of these networks are limited in terms of space-bandwidth product, which is the same as the number of waveguides. This essentially comes from the lack of dimensionality, as in integrated photonics, we effectively have one spatial dimension, making the space-bandwidth product to be $N \sim A/\lambda$, $A$ being the dimension of the chip and $\lambda$ being the optical wavelength. While the number of waveguides currently is much smaller due to technical limitations, integrated photonics is fundamentally limited by the achievable space-bandwidth product. Even with wavelength division multiplexing (WDM), the dimension of input vector remains small. As such, WDM can only provide a linear scaling of $N$. This limited space-bandwidth product necessitates time-domain multiplexing to send the data in batches, which also requires combining the data in the backend. As such most of these works did not report any excess electrical power and latency originating from the control circuit for the multiplexing. Moreover, some of the solved problems do not have an electronic benchmark and thus it is unclear if any advantage in the system level is achieved.

Most of the works essentially demonstrated a similar classification accuracy against a digital ANN, but the power and latency of an application specific digital IC were not calculated.

***Free space + Digital:*** Using free-space optics (either spatial light modulator or digital micro-mirror devices), researchers can achieve much larger-space bandwidth product [8]–[14]. Thanks to the two dimensions, the space-bandwidth product scales as $N \sim \left(\frac{A}{\lambda}\right)^2$, $A$ being the aperture of the optics and $\lambda$ being the optical wavelength. This is a much more favorable scaling than the linear scaling in integrated photonics (even with WDM). But most of the reported works neglect the power and latency coming from the conversion between the optics and electronics. Additionally, the use of a spatial light modulator can add substantial amount of power. Comparing these power numbers with a GPU is also unfair, as the GPU is designed to be a solution for many different problems, and thus have large redundancy. For a given problem, one can optimize and design an application specific integrated circuit, with pruning/ XNOR-operations and can require much lower power.

***All-optical:*** Finally, there are a few demonstrations (both free-space and integrated photonics [3], [8], [10], [15]) of all-optical neural network, which can be implemented without a digital backend. Some of them are fully linear (without any nonlinearity) and as such they cannot be used for complicated datasets. Moreover, even there, multiple layers of optics are generally used, each with additional scattering losses. Thus, the input power might be high, and they also require a laser to encode the electronic information. It is unclear what kind of wall-plug efficiency has been achieved in these cases. Thus, the true power is not reported. Others employed nonlinearity either based on saturable absorption, electromagnetic induced transparency in cold atoms, or optoelectronic nonlinearity (image intensifier or optically induced electro-optic nonlinearity). While some of them did demonstrate operation with MNIST data set, the power consumption of all the possible sources is not accounted for, for example, no laser powers are accounted for to cool down the atom. As such, a true estimation of system level power/ latency is missing in all these works, and thus the claims of benefit over purely digital ANN were not established.